\documentstyle[12pt]{article} 
\begin{document} 
\centerline{\LARGE { Similarity Analysis of Nonlinear Equations}}
\centerline{ \LARGE {and  Bases of Finite Wavelength Solitons}}
\vskip 1cm 
\centerline{A. Ludu, G. Stoitcheva and J. P. Draayer} 
\centerline{Department of Physics and Astronomy,} 
\centerline{Louisiana State University, Baton Rouge, LA 70803-4001} 

\begin{abstract}
We introduce a generalized similarity analysis which grants a qualitative description
of the localised solutions of any nonlinear differential equation.  This procedure provides
relations between amplitude, width, and velocity of the solutions, and it is shown to be 
useful in analysing  nonlinear structures like  solitons, dublets, triplets,  compact supported
solitons and other patterns. We also introduce  kink-antikink compact solutions  for a
nonlinear-nonlinear dispersion equation, and we construct a basis of finite wavelength
functions having self-similar properties. 
\end{abstract}

\vskip 3cm
\section{Introduction}

Nonlinear dynamics, which is a relatively new field of study, is a very important frontier
for probing natural phenomena. Active research efforts that focus on the mathematics and
physics of nonlinear dynamical systems have emerged worldwide in various fields,
including fluid dynamics, plasma physics, astrophysics, and even string theory \cite{general1}.
Among its unique features is the ability to describe a variety of patterns \cite{chaos} and
particle-like traveling solutions \cite{general2}. Other notable features of the theory include
its description of solitons and breather modes, features in quantum optics \cite{optics},
molecular and solid state physics phenomena \cite{molecular}, and  solitons in 
nuclear and particle physics \cite{nuclear}.

In contrast with linear theories which exhibit smooth regular motion, nonlinear models 
require nonlinear partial differential equations (NPDE) and  show  strong couplings between
different mechanisms and parts of the system. Also, the nonlinear interactions  involve
multiple scales
\cite{multiple}  and are related with self-similar  patterns or fractals. The NPDE solutions of
physical interest are mostly localized and demonstrate good stability in time and through
scattering with each other. Their shapes are related to the velocity,
thus making the nonlinear patterns distinct from linear waves. In the asymptotic domain
these solutions consist of isolated traveling pulses that are free of interactions. Close to
the scattering domain, the nonlinear solutions obey nonlinear superposition principles. 

The main challenge in any nonlinear analysis is the construction of localized or
finitely supported analytical solutions for the NPDE of
interest. This challenge includes issues regarding the inexistence of a superposition principle
for such solutions. Recent examples show that the traditional nonlinear tools (inverse
scattering, group symmetry, functional transforms) are not always applicable
\cite{compacton1,prl}. On the other hand, from an experimental  point of view one knows that
patterns which are observed in nature -- either stationary, growing, or propagating --
generally have finite space-time extension and a multi-scale structure. Since soliton, even
when they are localized, have an infinite extent, one needs other appropriate structures,
and eventually self-similar bases. 

In this paper, wavelet-inspired approaches for localized solutions of NPDE are explored.
We propose a new similarity formalism for the qualitative analysis, and
clasiffication, of soliton solutions of nonlinear equations. This method provides
relations between the characteristics of such solutions (amplitude, width and velocity) without
the need of solving the corresponding NPDE. The method uses the
multi-resolution analysis
\cite{wavelet1} where traditional tools like the Fourier integrals or linear harmonic analysis
are inadequate for describing the system. Wavelets are functions that have a
space-dependent scale which renders them an invaluable tool for analyzing multi-scale
phenomena. Wavelets have been used in signal processing, in problems
involving singular potentials, pattern recognition, image compression, turbulence and even in
radar and acoustic problems \cite{wavelet1,wavelet2}. Moreover, the
introduction of wavelet analysis in a study of NPDE is very natural because they can acommodate
everything from strong variations, even singularities, to a smooth behavior.

The second purpose of the paper is to show an example of construction of a nonlinear basis
for a NPDE with nonlinear dispersion.  There are many
physical reasons favoring wavelets in the construction of nonlinear bases. For example, the
breakup process of fluid drops has been shown to be self-similar, and in particular, the
singularities (necks) look identical at any scale
\cite{bona}.  Other nonlinear oscillations of  liquid drops,  shells, bubbles, or even neutron
stars involve such types of behavior \cite{chaos,prl}. 
We introduce in section 2 a qualitative similarity analysis for NPDE that yields
relations between the amplitude, width and velocity of their traveling solutions, with many 
examples and predictions. In section 3, we introduce a nonlinear wavelet-like frame, associated
with localized analytic solutions of a modified KdV equation, with nonlinear dispresion.

\section{Qualitative similarity analysis}

Finding analytic solutions  for the nonlinear equations which describe physical phenomena is
rather an exception than the rule. The traditional nonlinear tools like the inverse
scattering theory or functional transforms are not always applicable \cite{compacton1,fred}.
Also, when  phenomena of interest have many space (or time) scales, hundred 
times less or smaller than the  dimension (or time scale) of the whole system,
numerical methods may fail. This is the case of sharp propagating perturbations
developing discontinuities, or shock waves. A simple option is the expansion of
solutions in a basis of appropriate chosen linear modes. The most common choice are the  Fourier
series which have the advantage of orthogonality, but can hardly discriminate local behavior of
phenomena.  Moreover, information about the order of magnitude of the Fourier coefficients of a
signal or wave $u(x,t)$ is not sufficient for making conclusions about the size or scale of $u$.

In these special situations, the use of bases of multiresolution analysis and  wavelets has
become popular \cite{wavelet3}. We introduce a qualitative analysis for the localized 
traveling solutions belonging to any type
of NPDE, in terms of Morlet continuous wavelet approach. 
In our  case of interest, traveling localized perturbations,
the most  important information are provided  by the scale of the pulse (or the width) denoted
$L$, the amplitude $A$, and the group velocity $V$. The qualitative analysis in this section
provides simple relations between these three parameters, without actually solving the
equation. It also gives an estimation about the specific scale of the solutions.

The procedure consistsin the  substitution of all the following terms in the NPDE, according to
the rule 
\begin{equation}
u_t \rightarrow \pm Vu_x ,  \ \ 
u \rightarrow \pm A,  \ \ \
u_x \rightarrow \pm A/L,  \ \ \ 
u_{xx} \rightarrow \pm A/L^2 \dots ,
\end{equation}
and so forth for higher order of derivatives.
Consequently, the NPDE is mapped into an algebraic equation in $A,L$
and $V$. In Table 1 we present several examples of application of this substitution to
some well-known and widely used NPDE in physics. 

The validity of the method follows from the expansion of the soliton-like 
solution $u(x)$ in Morlet wavelets \cite{wavelet1}
\begin{equation}
\Psi _{\alpha }(x)=\pi ^{-1/4} e^{-i\alpha x-{{x^2}\over 2}},
\end{equation}
where $\alpha$ describes the scale of this mother wavelet. The support of any
Morlet wavelet is mainly confined in the $(-1,1)$ interval. We have the discrete Morlet wavelet
expansion of $u$
$$
u(x)=\sum_{j}\sum_{k}C_{j,k}\Psi _{\alpha } (2^j x -k),
$$
in terms of integer translations and dyadic dilations of the mother wavelet.
In order to reduce the number of scales needed, the range of the summation
should be choosen with an eye to the underlying physics.
We use in the following the asymptotic formula describing the pointwise
behavior of the Morlet wavelet series around of a point $x_0$ of interest \cite{approx}.
For a chosen $x_0$ and scale $j$, there is only one $k$ and $|\epsilon | \leq 1$
such that the support of the corresponding $\Psi _{\alpha, j,k}$ contains this point,
$k=2^j x_0 +\epsilon$. We can express the solution and its derivatives in a neighborhood of this
point 
$$
u(x_0) \approx \Psi (-\epsilon )\sum_{j}C_{j,2^j x_0 +\epsilon}\equiv \sum_j u_j (x_0 ) ,
$$
\begin{equation}
{{du} \over {dx}}(x_0) \approx -i \Psi (-\epsilon )\sum_{j}2^j  \alpha C_{j,2^j
x_0 +\epsilon}=-i\alpha \sum_j 2^j u_j (x_0 )  ,
\end{equation}
for $\alpha $ chosen enough large compared to $\epsilon $. Since the coefficient $1/ \alpha 2^j
$ represents the scale for each $\Psi _{\alpha , j, k}$ Morlet wavelet, we can define it as
a characteristic half-width
$L_{j}$. And we finaly have in $x_0$, from eqs.(3)
\begin{equation}
{{d^n u} \over {dx^n}}(x_0 )\approx \sum_{j} {{u_{j}(x_0 )}\over {L_{j}^{n}}},
\end{equation}
where $u_j (x_0 )=\Psi _{\alpha }(-\epsilon ) C_{j,2^j x_0 + 
\epsilon}\approx \Psi _{\alpha }(0) C_{j,2^jx_0 }$. Eq.(4) is the many scales generalization of
the simpler formula in eq(1). With eq.(4) in hand we can investigate the structure of
hypothetic soliton solutions, by choosing $x_0$ in the neighborhood of the maximum value of the
solution,  $u(x_0, 0)=A$.  Around this maximum, such solutions can be described very well by a
unique scale $L$, and hence the solution and its derivatives can be approximated  with the
corresponding dominant term, by  the substitutions in eqs.(1).

In Table 1 we present a series of examples of  NPDE, identified in the first column by the name
and the form of the equation. In the second column, we write the corresponding traveling
localized solution, if an analytical form is known. Such  solutions provide special
relations between $L,A$ and $V$, which are given in the third column. In the last column we
introduce for comparison the results of this Morlet  qualitative analysis, that is the 
relations between the three parameters, provided  by  eqs.(1).   The
usefulness of the approach may be checked, by a quick comparison between the fourth and the fith
columns. While the results in the fourth column are possible only when one knows
the analytical solutions, the results presented in the last column,  obtained by the
similarity approach, result directly from the NPDE, without actually solving it.

The first line in the Table presents a linear case, for comparative
purposes. The above Morlet wavelet approximation provides a correct expression 
for the dispersion relation ($V=c$ $\rightarrow$ $k^2=\omega^2/c^2$) with no constraint on
either the amplitude $A$ or on the width $L$.

The case of the KdV equation is described in the second row of the Table 1. The 
method gives a general expression for $L=L(A,V)$. For $L$ to be  related
to $A$ only, from the fourth column it results that the velocity $V$ must be proportional 
to $A$. In this case we obtain exactly  the well-known relation (column three) among
the parameters in the exact solution. A prediction of the method is if we allow
$V$ to depend on a power of $A$. This means solutions with a higher nonlinear coupling between
the shape and kinematics. A side effect would be a lower limit for $A$. Smaller solitons
than this limit can move vith velocity proportional to the amplitude only. 

The
same result is obtained for the MKdV equation (third row), except that in this
case $V$ needs to be proportional to the square of $A$ in order to have $L$ a
simple function of $A$ only. This prediction is again identical with that in the 
exact solution (third column).  Moreover, the same relations remain valid even for 
the new  exotic
solutions of the MKdV equation of compacton type,  \cite{electronic}
$$
u(x,t)={{\sqrt{32}k \cos [k(x-4k^2 t)]^2 }\over {3(1-{2\over 3} \cos [k(x-4k^2 t)]^2)}},
$$
which has $L= 5 \pi /6 k$ or $ \pi /6 k$, that is $L\sim 1/A$ like in the Table 1.

Next example (4-th row) is provided by a generalised KdV equation, in which the dispersion
term is quadratic
\begin{equation}
\eta _t + (\eta ^2)_x +(\eta ^2)_{xxx}=0.
\end{equation}
Eq.(5), known as K(2,2) from the two quadratic terms, admits compact supported
traveling solutions, named compactons
\cite{compacton1,fred,rosenau1,rosenau2}.  The compactons are powers of  trigonometric functions
defined on a half-period, and zero otherwise. In general, they have the form $Acos^{a}d(x-ct)$,
and different from solitons, their width is independent of the amplitude. 
This is the fact that provides a connection
with wavelet bases. They are characterize by a unique scale, and it is this feature
that makes it possible to introduce a nonlinear basis starting from this ``mother"
function.  For eq.(5) the compacton solution is given by
\begin{eqnarray}
\eta _c (x-Vt)={{4V}\over {3}} cos ^2 \biggl [  {{x-Vt}\over {4}} \biggr ], 
\end{eqnarray}
if $|x-Vt|<2\pi $ and zero otherwise. Here the velocity is a function of the amplitude. 
Notice that the width $L=4$ of the wave is independent of the
amplitude.  The quadratic
dispersion term is characteristic for the nonlinear coupling in a chain.  

The general compacton solution for eq.(5) is  actually a ''dilated" version of eq.(6).
That is, a combination of the first rising half of the squared cos in eq.(6), followed
by a flat domain of arbitrary length ($\lambda $), and finaly followed by the second,
descending part of eq.(6). Actually, this combination is just a kink compacton joined
smoothly with an antikink one
\begin{equation}
{\eta}_{kak} (x-Vt; \lambda)=
\left\{ \begin{array}{ll}
0 ...\\
{{4V}\over 3} \  cos^{2}\biggl [ {{x-Vt}\over 4} \biggr ],
\ \   -2\pi \leq x-Vt \leq 0
\\
{{4V}\over 3} ,  
\ \   0 \leq x-Vt \leq \lambda  \\ 
{{4V}\over 3} \  cos^{2}\biggl [ {{x-Vt-\lambda }\over 4} \biggr ],
\    \lambda \leq x -Vt \leq \lambda +2\pi 
\\ 
0 ... \\
\end{array}
\right.
\end{equation}
In Fig. 1 we present a  compacton, eq.(6),  a  kink-antikink
pair (KAK) described by eq.(7), both with the same amplitude and velocity. Although the second
derivative of this generalized compacton is discontinuous at its edges, the KAK, eq.(7),  is
still a solution of eq.(5) because the third derivative acts on $u^2$, which is a function of
class $C_3$. Finally, we can construct solutions by placing a compacton on the top of a KAK,
like in the third solution in Fig. 1. Such a solution exists only for a short interval of
time, since the two structures have different velocities. The solution is given by
\begin{equation}
\eta (x,t)=\eta _{kak} (x-Vt;\lambda )+\biggl ( \eta _c (x-V't-2\pi) + {{4V}\over 3 }
\biggr ) \chi ({{x-V't-2\pi} \over {2\pi}} ),
\end{equation}
for $0< t < (\lambda -4\pi)/(V'-V)$. Here $\chi(x)$ is the support function, equal with 1 for
$|x|\leq 1$ and 0 in the rest, and $V'=3\hbox{max}\{ \eta _c\}/4 +2V $.

For the K(2,2) compacton, eq.(6), the exact relations between the parameters are
$A=4V/3$ and $L=4$, \cite{rosenau1}. The relation provided by the similarity method, in the last
column of the forth row, predicts the existence of the compacton. That is, for a linear
dependence between the  amplitude and the speed, the half-width is constant and does not depend
on $A$ or
$V$. This fact ($L\equiv L_0=const.$) is a typical feature of  K(2,2) compactons. Moreover, in
literature there was found numerically that for any compact supported  initial data, widder
than $L_0$, the solution decomposes in time into a series of
$L_0$ compactons, Fig. 2. For narrower initial data the numeric solution blows up. There
is no exact or analytic explanation of this effect, so far.  The similarity method can give a
hint in this situation, too, by using the graphic of the relation
$L=L(V,A)$ provided by this qualitative method. In Fig. 3 $L$ is ploted
versus $V$, for several values of $A$ (larger values of $A$ translate the curves to the right).
The  half-width of a stable compacton was chosen $L_{0}=0.707$. Above this value, Fig. 3a, 
wider compact pulses produce an intersection for each curve (each $A$) with the axis $L_0$
providing series of compactons of different heights, like in the numerical experiments,
\cite{prl,rosenau1}. Below this $L_0$ line, all the curves approache infinite amplitude,
providing  instability of narrower shapes.

Another good example  of prediction of the method is exemplified 
in the case of a general convection-nonlinear dispersion
equations, denoted K(n,m)
\begin{equation}
\eta _t + (\eta ^n)_x +(\eta ^m)_{xxx}=0.
\end{equation}
Compacton solution for any $n\neq m$ are not known in general, except some particular cases. In
this case we find a general relation among the parametrs, for any $n,m$, shown in  the
$5^{th}$ and $6^{th}$ rows. These general relations $L(A,V)$  approache the known relations for
the  exact solutions, in the  particular cases like $n=m$ ($5^{th}$ row),
$n=m=2$ ($4^{th}$ row), $n=m=3$. And $n=3,m=2$;
$n=2,m=3$ in the $6^{th}$ row.  These results can be used to predict the behavior of
solutions for all values of $n,m$.

Similar analysis can be done in the case of sine-Gordon equation, if we ask that velocity
be proportional with $L^2$ ($7^{th}$ row). In this case we obtaining a transcendental equation
in $A$, which is just the case of the  sine-Gordon soliton.
In the $8^{th}$ row, we present the cubic nonlinear Schr\"odinger equation (NLS) which has a
soliton solution, too \cite{nuclear}. This equation arises, for example, in nonlinear optics or
in the polaron model in solid state physics, \cite{optics, molecular}. In the general case of a
NLS of order
$n$ ($9^{th}$ row), when the general analytical solution is unknown, the method predicts a
special
$L=L(A,V)$ dependence shown in the fourth column and in Fig. 4. Contrary to third order NLS,
where the dependence of $L$ with $A$ is monotonous for $V=\sim \pm A$ ($n=3$ in Fig. 4), at
higher order, the $L(A)$ function has a discontinuity in the first derivative. This wigle of the
function (Fig. 4 for $n=4$) yields at a  critical width, producing
bifurcations in the solutions and scales. As a consequence,  initial data close to this
width can split into  doublet (or even triplet for higher order NLS) solutions, with different
amplitudes. Such phenomena have been put into evidence in several numerical experiments for
quintic nonlinear equations \cite{rosenau1,rosenau2}.

In the following, we present  another example of applications of this qualitative approach,
related to a new type of behavior of nonlinear systems. Traditional solitons  move
with constant speed on a rectilinear path (except for the  roton, \cite{prl} which has a
circular trajectory with constant angular velocity). The speed is usually equal with the
amplitude scaled with a constant. Higher solitons travel faster and there are no 
solitons at rest (zero speed asks for zero amplitude). They can travel in  both
directions with oposite signs for the amplitude.
The situation is  different in the case of compactons, which allow also stationary
solutions. When linear and nonlinear disspersion occur simultaneously, like in  the so
called K(2,1,2) equation
$$
u_t +(u^2 )_{x} + (u)_{xxx} +\epsilon (u^2 )_{xxx}=0,
$$
where $\epsilon $ is a control parameter, the similarity approach yields a dependence of the
form  
$$
L=\sqrt{(\pm A+\epsilon )/(V\pm A)},
$$
which still provides a constant width if $V=\pm A +2\epsilon $.  In this case the speed is
proportional with the amplitude, but can change its sign even at non-zero amplitude.
Solutions with larger amplitude than a critical one ($A_{crit}=\mp 2\epsilon$)
move to the right, solutions having the  critical amplitude are at rest, and solutions
smaller than the critical amplitude move to the left. This behavior was 
explored in \cite{rosenau1}. However, such a switching of the speed is
not necessarily a feature of the nonlinear dispersion. 
A  compacton of amplitude $A$ on the top of a KAK solution of amplitude $\delta $
\begin{equation}
u(x,t)= A cos ^2 \biggl ( {{x-Vt}\over{4}}  \biggr ) +\delta , 
\end{equation}
is still  a solution of the  K(2,2) equation, $u_t +(u^2 )_x +(u^2 )_{xxx}=0$, with the
velocity given by $V={3 \over 4}\biggl (2 \delta + A  \biggr )$. For $A=-2 \delta$ the
compacton becomes a  stationary anticompacton, embedded in the moving, supporting  KAK. Such an
example is presented in Fig. 5 for a slow-scale time-dependent amplitude compacton. The 
induced oscillations in the amplitude transform into oscillations in the velocity.
While not the topic of this paper, such a dynamic system has been analysed and it will be
published soon. The key to such a conversion of oscillations is the coupling between the
traditional nonlinear picture (convection-dispersion-diffusion) and the typical
Schr\"odinger terms. 

A last application of this method, occurs if the KdV equation has an
additional term depending on the square of the curvature
\begin{equation}
u_t +uu_x +u_{xxx} +\epsilon (u_{xx}^{2})_{x}=0.
\end{equation}
This is the case for extremely sharp surfaces (surface waves in solids or
granular materials) when the hydrodynamic surface pressure cannot be linearized 
in curvature. Such a new term yields a new type of localized solution fulfilling the relations
$$
L=\sqrt{
{{4\epsilon A}\over {\pm \sqrt{1-8\epsilon  A(A\pm V)}-1}}}.
$$
If we look for a constant half-width solution (compacton of $1/L =\alpha $) we need
a dependence of velocity of the form $V=(1+\alpha ^2 \epsilon /8)A+1/8\epsilon A +\alpha /4$.
There are many new effects in this situation. The non-monoton dependence of the
speed on $A$ introduces again bifurcations of a unique pulse in dublets and triplets.
Also, there is  a upper bound for the amplitude at some critical values of the width. Pulses
narrower than this critical width drop to zero. Such bumps can exist in
pairs of identical amplitude at different widths. They may be related with
the recent observed "oscillations" in granular materials, \cite{rosenau1,oscillon}.

As  the examples presented in Table 1 proved  the above method provides a reliable
criterium for finding compact suported solutions. The reason this simple
prescription works in so many cases follows from the advantages of wavelet
analysis on localized solutions. We stress that this method has little to do
with the traditional similarity (dimensional) analysis
\cite{compacton1,bona,fred,rosenau1,rosenau2,simil}. In the latter case one obtains
relations among powers of $A, L$ and $V$, not relations with numeric coefficients like those
found in our method.

\section{Compacton  kink-antikink pairs and the multiresolution frame}

A common feature of all NPDE and of the finite
differences equations is the existence of compact supported solutions. Compactons and discrete
wavelets are typical examples.  An interesting general conclusion can be obtained if we look
at a one-dimensional model described by the most general NPDE dynamical equation 
\begin{equation}
\partial _t u ={\cal O}(x,{\partial }_{x} ) u, \label{general dynamic}
\end{equation}
where ${\cal O}$ is a nonlinear differential operator. By taking into account {\it
only} traveling solutions, this NPDE reduces to a NODE in the 
coordinate $\xi =x-Vt$ for an arbitrary velocity
$V$. If $u(\xi )$ is a compact  solution it results that it
is not unique for given initial compact data. If one chooses  zero initial value for the
solution and its derivatives up to the requested, in a certain point $\xi
_0$ of the $\xi $ axis, these conditions can be fulfilled by any
linear combination of disjoint translated versions of one particular solution, placed everywhere
on the axis except  $\xi _0$. Consequently, for such initial data, the solution is not unique.
This result shows that the compact supported property of the initial data and the solution,
implies its non-uniqueness.

Since we can transform the NODE into a nonlinear differential
system of order one
\begin{equation}
{{d{\vec U}}\over {dx}} ={\vec F}(\xi ,
 {\vec U}), \ \ \ {\vec U}=(u,\partial _x u, ...) , 
\end{equation}
we can apply the fundamental theorem of existence and uniqueness to solutions
of eq.(13), for  given initial data ${\vec U}(\xi _0 )={\vec U}_0$.
If the function ${\vec F}$ in eq.(13) fulfills the Lipschitz condition  (its relative
variation is bounded) than, for any initial condition, the solution is unique, \cite{hure}.
Since  any linear function is analytic and hence Lipschitz, we  conclude
that only nonlinear functions ${\vec F}$ allow the existence of
compact supported solutions. Thus, a compact soliton 
implies non-uniqueness in the underlying NPDE, which implies non-Lipschitzian structure
of the NPDE and hence the existence of nonlinear terms.

In the following we investigate some compact solutions of the K(2,2) equation.
The high stability against scattering of the K(2,2) compactons,   or 
compacton generation from compact initial data, suggest  they may play the role of a nonlinear
local basis. We know form many numerical experiments, \cite{compacton1,fred,rosenau1,rosenau2},
that  any positive compact initial data  decomposes into finite series of compactons and
anticompactons. This suggests that an intrinsic ingredient for  a nonlinear basis could be the
multiresolution structure of the solutions, similar with the structure of scaling functions in
wavelet theory.

The compactons given in eqs.(6,7) have constant half-width  and hence
describes a unique scale, which can cover all the space by integer translations. From the point
of view of multi-resolution analysis, the K(2,2) equations acts like a
$L$-band filter, allowing only a particular scale to emerge for any given set
of initial condition. To each scale, from zero to infinity, we can associate a K(2,2)
equation with different coefficients. However, the compacton solution is not the
unique one with this property.  For a given K(2,2) equation, we can thus extend the scale from
$L$  to any larger scale. These more general compact supported solutions are still
$C_{2}({\bf R})$ and are combinations of piece-wise constant and piece-wise $\cos ^2$
functions. The simplest shape is given by a half-compacton prolonged with a constant level,
that is a kink solution. The
basis solution is a kink-antikink (KAK) compact supported combination, Fig. 1. Such
kink-antikink pairs of different length, can be associated with other compactons, or KAK pairs,
one on the top of the other
\begin{equation}
{\eta}_{comp+KAK} (x-Vt; \lambda)=
\left\{ \begin{array}{ll}
0 ...\\
{{4V}\over 3} \  cos^{2}\biggl [ {{x-Vt}\over 4} \biggr ],
\ \   -2\pi \leq x-Vt \leq 0 \\
{{4V}\over 3} ,  \ \   0 \leq x-Vt \leq \delta  \\
{{4V}\over 3}+{{4}\over 3}(V'-2V) \  cos^{2}\biggl [ {{x-V't}\over 4} \biggr ],
\ \   \delta \leq x-Vt \leq \delta +4\pi \\
{{4V}\over 3} ,  \ \   \delta+4\pi  \leq x-Vt \leq \lambda \\ 
{{4V}\over 3} \  cos^{2}\biggl [ {{x-Vt-\lambda }\over 4} \biggr ],
\    \lambda \leq x -Vt \leq \lambda +2\pi 
\\ 
0 ... \\
\end{array}
\right.
\end{equation}
where $\delta < \lambda$ characterizes the initial position (at $t=0$) of the top compacton,
with respect to the flat part of the KAK solution. The amplitude $4(V'-2V)/3$ of the compacton,
and the amplitude $4V/3$ of the KAK,  are related to their velocities $V'$ and $V$,
respectively.  The length of the flat part, $\lambda $, is arbitrary. A compound solution is
not stable in time since the different elements travel with different velocities. The total
height of the compacton is $4(V'-V)/3$. Since the higher the amplitude is, the faster the
structure travels, the top compacton moves faster than the KAK, and at a certain moment it
passes the KAK. Because the area  of the solution is conserving, such a compound structure
decomposes into compactons and KAK pairs. Similar and even more complicated constructions can be
imagined, with indefinite number of compactons and KAK's,  if one just
fulfills the $C_3$ continuity condition for the square of the total structure.
Such structures, defined at the initial moment can interpolate any function, playing a similar
role with wavelets or spline bases. 
It has been also proved that the KAK solutions are stable, by using both
a linear stability analysis and  Lyapunov stability criteria, \cite{fred,simil}.

For a given K(2,2) equation, the compacton solution, eq.(6) and in addition the family of KAK
solutions, eq.(7) can be organized as a scaling functions system. They  act
like a low-pass filter in terms of space-time scales and 
give the opportunity to construct frames of functions from the wavelet model,
\cite{wavelet1,wavelet2,wavelet3}. 

For the sake of simplicity we will renormalize the coefficients of the
K(2,2) equation such that the support of the simple compacton is one.
That is, we take $\eta _c (x,t)=\eta _{kak}(\pi (x-Vt),0)$ on the interval $|x-Vt|$
in $[-1/2,1/2]$. We construct a multiresolution approximation of $L^{2}({\bf R})$, that is an
increasing sequence of closed subspaces $V_j$, $j\in {\bf Z}$, of $L^{2}({\bf
R})$ with the following properties, \cite{wavelet2,wavelet3}
\begin{enumerate}
\item
The $V_j$ subspaces are all disjoint and their union is dense in
$L^{2}({\bf R})$.

\item
For any function $f \in L^{2}({\bf R})$ and for any integer $j$ we have
$f(x) \in V_j $ if and only if $D^{-1}f(x) \in V_{j-1}$ where $D^{-1}$ is an
operator that will be defined later.

\item
For any function $f \in L^{2}({\bf R})$ and for any integer $k$, we have
$f(x) \in V_0 $ is equivalent with $f(x-k) \in V_0$.

\item
There is a function $g(x) \in V_0$ such that the sequence $ g(x-k)$  with
$k\in {\bf Z}$ is a Riesz basis of $V_0$.

\end{enumerate}
In the case of compact solutions of K(2,2) of unit length, we chose for the
space $V_0$ that which is generated by all translation of $\eta _c$ with any
integer $k$. The subspaces $V_j$ for $j \geq 0$ are generated by all integer
translations of the compressed version of this function, namely, by
$\eta_{kak}  (2^{j}\pi(x-Vt),0)$. The subspaces $V_j$ for $j\leq 0$
are generated by all integer translations of the KAK solution of length $\lambda 2^j -1$. For
example, $V_{-1} $ is generated by $\eta_{kak} (\pi (x-2^{j}Vt),0)$. The spaces $V_j, j\geq
0$ are all solutions of K(2,2); the others are not.  The function $g(x)$ is taken to be 
$\eta_{kak}
(\pi (x-Vt),0)$. It is not difficult to prove that these definitions fulfill restrictions
one, three, and four. As for the second criterion, we define the action of the operator 
$D^{-1} f(x) = f(2x)$ if $f(x) \in V_{j}$ with a $j$ positive integer, and
$D^{-1}\eta_{kak} (\pi2^{j}(x-2^{j}Vt),2^{-j}-1)=\eta_{kak} (\pi 2^{j}(x-2^{-j+1}Vt), 2^{-j+1}-1
)$ for negative
$j$. In conclusion, we construct a frame of functions made of contractions of compactons
 and sequences of KAK solutions. We can write the corresponding two-scale
equation which connects the subspaces (the equivalent of eq.(15)),
\begin{equation}
\eta_{kak} (\pi (x-Vt), 1)=\eta_{kak} (\pi (x-Vt),0)+\eta_{kak} (\pi (x-Vt-1),0).
\end{equation}
We will denote generically by $\eta _{k,j}$ the elements of this frame, that is
$$
\eta _{k,j} (x)=\eta_{kak} (\pi (x-2^{j}Vt-k), 2^{j}-1)|_{t=0},
$$
where $t=0$ means that we neglect the time evolution, but the amplitude is still amplified  with
a factor of $2^j$, in virtue of relation $\eta_{max}=4V/3$. In the following, we can expand any
initial data for the K(2,2) equation in this basis.
\begin{equation}
u_{0} (x) = \sum_{k}\sum _{j} C_{k,j} \eta _{k,j} (x).
\end{equation}
We notice that the following equality holds for $j'$,$j$
\begin{equation}
\eta _{k,j} \eta _{k',j'}
\left\{ \begin{array}{ll}
\neq 0 & k'=k\cdot 2^{j'-j}, ..., (k+1)\cdot 2^{j'-j}-1 \\
=0 & \mbox{otherwise.}
\end{array}
\right.
\end{equation}
After some rather elaborate algebraic calculations and by using eq.(21), we show
that the square of this function (since the equations is nonlinear and of order
two) will be given by
$$
u^2 (x)=\sum _{k,j} \sum _{j' \geq j} \sum _{k' \in I}C_{k,j}C_{k',j'}
$$
\begin{equation}
\times \biggl (
\sum_{i_1 =0}^{1} \sum_{i_2  =0}^{1}... \sum_{i_{j'-j}=0}^{1} \eta _{\sigma (i_1
,i_2 , ..., i_{j'-j}),j'}
\biggr ) \eta _{k',j'},
\end{equation}
where $I$ is the range of $k'$ described in the first line of eq. (21), and 
$$
\sigma (i_1 , i_2 , ..., i_{j'-j}) =
\sum_{l=1}^{j'-j} i_l 2^{j'-j'l+{{(j'-j)(j'-j+1)-l(l+1)}\over 2}}
$$
$$
+k2^{(j'-j)j+{{(j'-j)(j'-j+1)}\over 2}}.
$$
From eq.(21) we notice that in eq.(22)the unique nonzero terms are
those for which $\sigma (i_1 , i_2 ,..., i_{j'-j} )=k'$ with $k' \in I$. This result
express the following simple fact. The initial data is expanded in different
scales and different translations. The translations are mutually orthogonal so
they do not give a contribution to the square. When we have to multiply two
different scales in the expression of the square, we reduce the wider scale in
terms of linear combination of the narrower one by using the two-scale equation,
eq.(19). This is what eq.(22) expresses. Out of all the terms in such a product
only approximately  $(2^{-j}-1)/(2^{-j'}-1)\simeq 2^{j'-j}$ give non-zero
contributions. In other words this number is given by the number of solutions
of equation $\sigma (i_1 , i_2 ,..., i_{j'-j} )=k'$, with $k' \in I$. This is
the primary advantage of treating nonlinear problems with a basis that has a
scale criterion. Another advantage is that all the function in the basis are
actually contractions or dilations, and translations of only two basic ones.

\section{Comments and conclusions}

In the present paper  we introduce new applications for
wavelets, in the field of the study of localized solutions of  nonlinear differential
equations. The existence of compactons and discrete
wavelets underlines a common feature of NPDE and finite
differences equations, that  is the existence of compact supported solutions.
We propose a new similarity formalism for the qualitative analysis and clasiffication
of soliton solutions, without the need of solving the corresponding NPDE.
Also, we proved that starting from any unique soliton solution of a NPDE, we can construct a
frame of solutions organized  under a multiresolution criterium.
This approach provides the possibility of constructing
nonlinear basis  for NPDE. We show that frames
of self-similar functions are related with 
solitons with compact support.   In addition, we notice the evidence
that compactons fulfil both characteristics of solitons
and wavelets, suggesting possible new applications.
Such unifying direction between nonlinearity and self-similarity, can bring
new applications of wavelets in cluster formation, at any scale, from
supernovae through fluid dynamics to atomic and nuclear systems. The similarity approach
can be applied with succes to the physics of droplets,
bubbles, traveling patterns, fragmentation, fission and
inertial fusion.

\vskip 1cm 
Supported by the U.S. National Science Foundation through a regular grant,
No. 9970769, and a Cooperative Agreement, No. EPS-9720652, that includes
matching from the Louisiana Board of Regents Support Fund.

\vfill
\eject

\vfill
\eject


\vskip 1cm
\centerline{Figure Captions}
\vskip 1cm
\begin{itemize}

\item
Fig. 1

A compacton and a kink-antikink pair solution (KAK) of the equation K(2,2), both having the
same amplitude, and hence velocity V. To the right, there is a smaller compacton  on the
top of  KAK. The upper compacton has higher speed, V'.

\item
Fig. 2

A finite series of K(2,2) compactons emerging from  initial compact data, with the width
larger than the compacton width.

\item
Figs. 3

The half-width $L$ versus velocity $V$ for the K(2,2) equation, for different amplitudes A.
Figure 3a shows widths larger than ${L_{compacton}=3/4}$, and Fig. 3b shows
narrower widths, $L<3/4$. Amplitude increases from left to right, in the range
0.01-0.85.

\item
Fig. 4

The half-width $L$ plot versus amplitude $V$, for the third (n=3) and forth (n=4) order NLS
equation, in two  $V=\pm A$ cases. From the figure one notes that the quartic NLS
equation yields bifurcations in the solutions.

\item
Fig. 5

The solution of the mixed  linear plus nonlinear-dispersion K(2,2) equation,
in the case of a solution with a slow oscillating shape.

\end{itemize}

\vfill
\eject
\pagestyle{empty}
\begin{table}[t]
\caption{Nonlinear equations, exact solutions  and Morlet similarity
analysis.\label{tab:exp}}
\vspace{0.2cm}
\begin{center}
\begin{tabular}{cccc}
\hline \\
{\bf Equation} &{\bf Solution} &{\bf Relations} &{\bf Wavelet} \\
\hline \\
Linear wave & $\sum C_k e^{i(kx\pm \omega t )}$ & $k^2 = \omega ^2 /c^2$ & 
$V=c$ \\
$u_{xx}-(1/ c^2) u_{tt}=0$ & && A,L   arbitrary  \\
\hline \\

KdV=K(2,1)  & $A~sech ^2 {{x-Vt} \over L}$ &  $L=\sqrt{{2 \over A}}$ & 
$L={1 \over {\sqrt{|\pm V\pm 6A|}}}$\\
$u_{t}+6uu_{x}+u_{xxx}=0$ & &  $V=2A$ & \\ 
\hline \\
MKdV=K(3,1) &   $A~sech  {{x-Vt} \over L}$ & $L=1/A$ & $L={1 \over \sqrt{|\pm V \pm 6A^2 |}}$ 
\\
$u_{t}+u^2u_{x}+u_{xxx}=0$ & &  $A=\sqrt{V}$  &\\
\hline \\
K(2,2) &  $A\cos ^2 {{x-Vt}\over L}$ &  $L=4$ & 
$L=\sqrt{{{8A}\over {|\pm V \pm 2A|}}}$, if \\ $u_{t}+(u^2)_{x}+(u^2
)_{xxx}=0$  &$|(x-Vt)/L|\leq \pi /2$ & $ V=3A/4$ &  $V=-3A/2$, $L=4 $\\
\hline \\
K(n,n)  & $\biggl [ A cos^{2}\biggl ( {{x-Vt} \over {L}}
\biggr ) \biggr ] ^{1 \over {n-1}}$ 
& $A={{2Vn}\over {n+1}} $ &
$L=\sqrt{{{n(n^2 +1)}\over{\pm \alpha \pm n}}}$ \\   $u_t +(u^n )_x
+(u^n )_{xxx}=0$ &\ if $|x-Vt| \leq {{2n\pi}\over{n-1}}$ & 
$L={{4n}\over {(n-1)}}$ &
if $V=\alpha A^{n-1}$ 
\\ 
&and 0 else&& \\
\hline \\
K(n,m)  & unknown 
 &  & 
$L=\sqrt{{{n(n^2 +1) A^{n-1}} \over {\pm V \pm mA^{m-1}}}}$
\\  
$u_t +(u^n )_x +(u^m )_{xxx}=0$ & if $n\neq m$ & 
 & \\ 
\hline \\
sine-Gordon  & $A~\tan^{-1}\gamma ~e^{{x-Vt}\over {L}}$ & $ A=4 $ & 
$\pm {{VA}\over {L^2 }}=sin
A$ \\   $u_{xt}-\sin u=0$ & &  $V=L^2$ & \\ 
\hline \\
NLS(3)  & $Ae^{i(\omega t + kx)} sech  {{x-Vt} \over L}$ &  $L={1 \over A}$ & 
$L={{\pm V \pm \sqrt{|V^2 - 4 A^2 |}}\over {2A^2 }}$\\
$\Psi_{t}+\Psi _{xx}+\Psi ^3=0$ & &  $A\simeq V$ & \\ 
\hline \\
NLS(n)  & unknown &   & 
$L={{\pm V \pm \sqrt{|V^2 - 4 A^n |}}\over {2A^n }}$\\
$\Psi_{t}+\Psi _{xx}+\Psi ^n=0$ & &  & \\ 
\hline
\end{tabular}
\end{center}
\end{table}

\end{document}